\documentclass[
    ,final            
  ]
  {aipproc}

\layoutstyle{6x9}

\begin{document}

\title{The Fluorescence Detector of the Pierre Auger Observatory - a Calorimeter for UHECR}


\author{B.~Keilhauer for the Pierre Auger Collaboration }{
  address={Universit\"at Karlsruhe, Institut f\"ur Experimentelle Kernphysik, 
   Postfach 3640, 76021 Karlsruhe, Germany and \\}
  {Observatorio Pierre Auger, Av.~San Martin Norte 304, 5613 Malarg\"ue,
     Argentina; http://www.auger.org/auger-authors.pdf}
}

\begin{abstract}
The Pierre Auger Observatory is a hybrid detector for ultrahigh energy cosmic rays 
(UHECR) with energies above 10$^{18.5}$~eV. Currently the first part of the Observatory 
nears completion in the southern hemisphere in Argentina. One detection technique uses 
over 1600 water Cherenkov tanks at ground where samples of secondary particles of extensive 
air showers (EAS) are detected. The second technique is a calorimetric measurement of 
the energy deposited by EAS in the atmosphere. Charged secondary particles of EAS lose
part of their energy in the atmosphere via ionization. The deposited energy is converted 
into excitation of molecules of the air and afterwards partly emitted as fluorescence light 
mainly from nitrogen in the wavelength region between 300 and 400~nm. This light is 
observed with 24 fluorescence telescopes in 4 stations placed at the boundary of the 
surface array. This setup provides a combined measurement of the longitudinal 
shower development and the lateral particle distribution at ground of the same 
event. Details on the fluorescence 
technique and the necessary atmospheric monitoring will be presented, as well 
as first physics results on UHECR.
\end{abstract}

\maketitle

\section{Introduction}

Ultrahigh energy cosmic rays (UHECR) are a continuous particle flux arriving at the 
Earth's atmosphere with strongly decreasing rate at highest energies. At energies 
above 10$^{18}$~eV, their origin is expected to be
extragalactic. However, even after several years of measurements, the major
questions remain unanswered: where and how do these particles
obtain such high energies and what is their elemental composition. For 
resolving the enigma about UHECR, experiments are
designed for measuring the primary energy and type of cosmic rays as well as their 
arrival directions.

Entering the Earth's atmosphere, cosmic rays initiate extensive air showers (EAS).
The primary particle of an EAS, which is the actual cosmic ray particle, interacts
with a nucleus of air molecules. This hadronic interaction yields 
new secondary particles which will interact themselves with nuclei of air 
molecules. A growing cascade of secondary particles develops in the Earth's
atmosphere, which is called an extensive air shower. After an initial phase
of particle multiplication, the number of charged particles in the shower
falls due to energy losses. For inclined showers with moderate primary energy, 
only some of the overwhelming amount of particles reaches the 
Earth's surface. These particles can be observed by particle detectors at ground.
For UHECR with nearly vertical incidence, the maximum of shower development
is reached close to ground or even ``below'' ground.

Another detection technique arises from the energy losses of EAS in the atmosphere.
The energy deposited in the air excites molecules of the atmosphere where
particularly the excitation of nitrogen molecules is of interest. The de-excitation
of nitrogen molecules is partly proceeding through the channel of fluorescence emission. 
So a light track of the air shower in the UV-wavelength region between 300 and
400 nm can be detected by adequate telescopes. This enables a 
calorimetric measurement of the deposited energy of EAS which can be
used to obtain the primary energy of the cosmic ray.

\section{The Pierre Auger Observatory}

The Pierre Auger Observatory is currently the only data-taking experiment
for the detection of UHECR but still under construction. The first part nears 
completion in the Pampa Amarilla, near Malarg\"ue, Argentina~\cite{auger_NIM}. The 
second part will be installed in Colorado, USA. Each site of the observatory
is designed as a hybrid detector, which consist of particle detectors at
ground and fluorescence telescopes. This configuration ensures at least
in dark nights the possibility of simultaneous detection of
air showers with two different detection techniques.

In May 2006, about 1100 of the total 1600 water Cherenkov tanks, as
particle detectors, have been deployed in the Pampa. The spacing of the
tanks is 1.5~km and finally an area of 3000~km$^2$ will be covered. Within
these detectors, energetic electrons and muons induce flashes of Cherenkov light
which are recorded by three photomultipliers inside the water-filled tank. The 
duty cycle of this system is nearly 100~\%~\cite{SD_ICRC}.

The second detector component is fluorescence telescopes. Four stations are 
located at the boundary of the surface array each consisting of six telescopes.
A single telescope has a field of view of roughly 30$^\circ \times 30^\circ$
with a minimum elevation of 1.5$^\circ$ above horizon. Each station in total has
180$^\circ$ in azimuth observing the atmosphere above the surface array.
The telescopes are built with a \emph{Schmidt} optics with an UV-filter
for reducing the background light. This system is operated
during the one week before and after new moon at night. Therefore, 
the duty cycle reaches only 15~\% of the surface detectors. The 
camera in the focal point consists of 440 photomultipliers each
covering 1.5$^\circ$~\cite{FD_ICRC}.

Data taking started in January 2004 with a continuously increasing
detector. Until June 2005, an exposure of 1750 km$^2$ sr yr could be
achieved~\cite{sommers}.

\section{Extensive Air Showers}

The development of extensive air showers in the atmosphere is strongly
fluctuating from shower to shower. Nevertheless, some systematic 
behavior is correlated with the energy and type of the primary particle.
High-energetic shower penetrate deeper in the atmosphere than lower-energetic
ones. Vertical showers develop deeper in the atmosphere than
more inclined ones. And also proton-induced air showers develop deeper 
than air showers induced by more heavy nuclei like iron. The shower 
development is described according to the amount of traversed air which
is the \emph{atmospheric depth} $X$ given by
\begin{eqnarray}
X(h_0) = \int_{h_0}^\infty \rho_{air}(h)~dh,
\end{eqnarray}
with $h$ as geometrical height and $\varrho_{air}$ as the altitude-dependent air 
density\footnote{Here written for a vertical particle trajectory.}.
 
As the density of air is quite low at higher altitudes, the height of first interaction is
strongly fluctuating. The incoming particles represent the projectiles and the
resting air nuclei are the target similar to the setup in fixed-target accelerator
experiments. In the case of nuclei, not all
nucleons of the projectile interact with the target, most of them are only
spectators~\cite{knapp}.

In general, the description of EAS can be divided into three particle components:
the electromagnetic, muonic, and hadronic. If the incoming particle is a nucleus, 
high energy hadronic interactions produce
mostly nucleons, charged and neutral pions, and kaons. The $\pi^\pm$ and $K$ 
mesons have relatively long lifetimes of 10$^{-8}$~s, thus they can
interact with air nuclei or decay depending on their energy \cite{khristiansen 1980}. If the
$\pi^\pm$ decay before interacting, they give rise to the muonic component accompanied by
neutrinos: $\pi^\pm \rightarrow \mu^\pm + \nu$. Many of the muons reach the Earth's
surface due to the relativistic time dilatation of their lifetime of 2.2 $\times
10^{-6}$~s and small energy loss. Those which decay, 
produce parts of the electromagnetic component:
$\mu^\pm \rightarrow e^\pm + \nu + \bar{\nu}$.

If the incoming particle is an 
electron or positron, energy loss due to Bremsstrahlung is one of the 
most dominant processes. The emitted high-energetic photons create new
pairs of electrons and positrons. In one radiation length, their energy is
attenuated by the factor $1/e$, which is approximately the same as the $1/e$ 
attenuation of a gamma-ray beam due to pair production. In air, the
radiation length $X_0^{air}$ is about 37~g/cm$^2$. The particle multiplication is large 
and the electromagnetic component is the
most numerous part in EAS. This electromagnetic part of an
air shower can also be induced by the decay of a $\pi^0$ into two
photons~\cite{sommers_CR}. Losing more and more energy, these particles
reach the critical energy which describes the equality of the energy loss due
to radiation and the loss due to ionization. Since ionization energy loss is about
2.2~MeV/g/cm$^2$, the critical energy in air is 81~MeV. The interplay of particle
production and energy loss yields in a shower maximum and closer to ground the shower
starts to die out. Only some of the electrons and positrons reach the Earth's 
surface.

\begin{figure}
\label{fig:longidev}
\begin{minipage}[c]{.50\linewidth}
  \includegraphics[width=.95\textwidth]{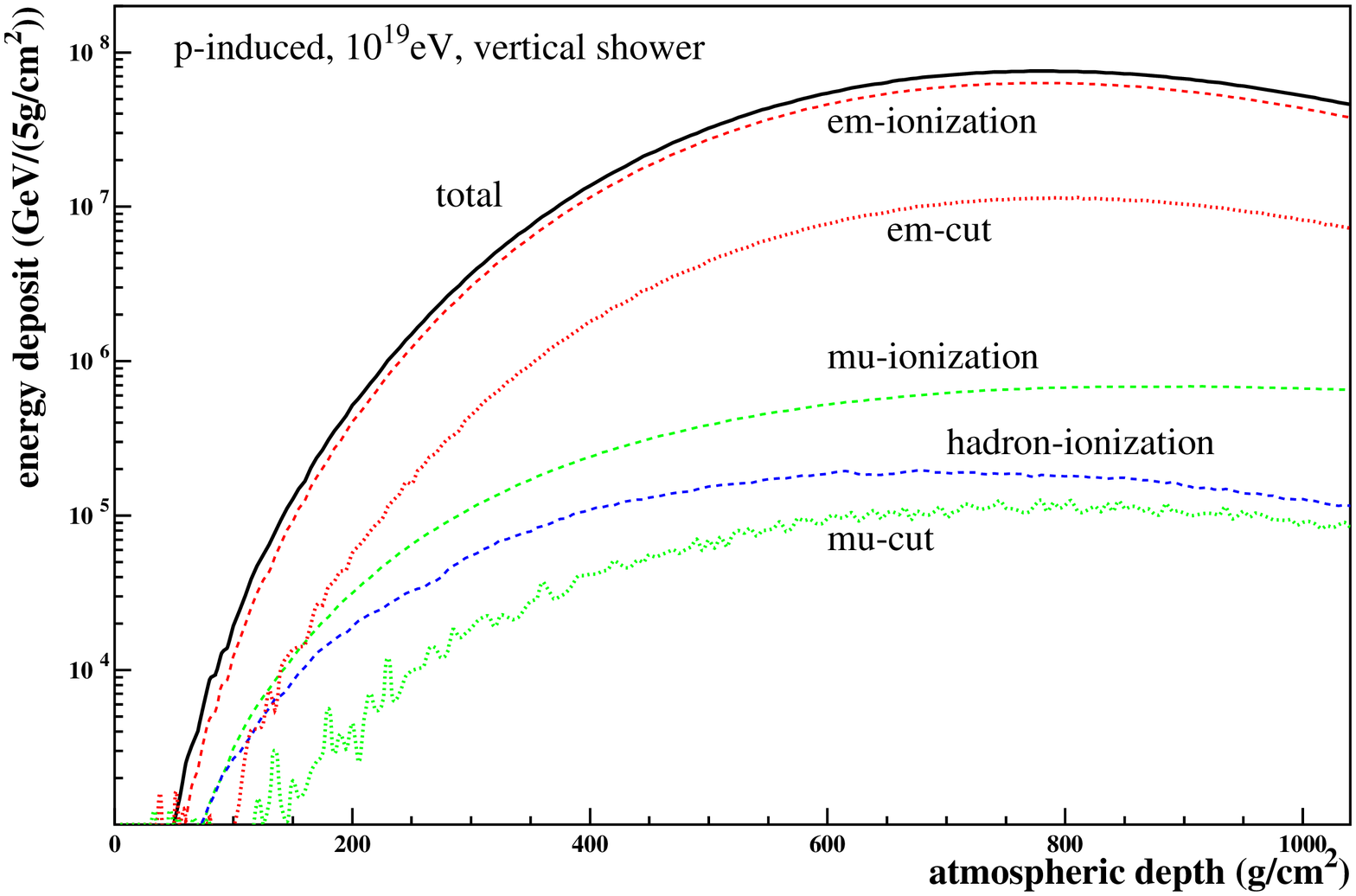}
  \caption{  }
\end{minipage}\hfill
\begin{minipage}[c]{.50\linewidth}
  \includegraphics[height=0.23\textheight,width=.95\textwidth]{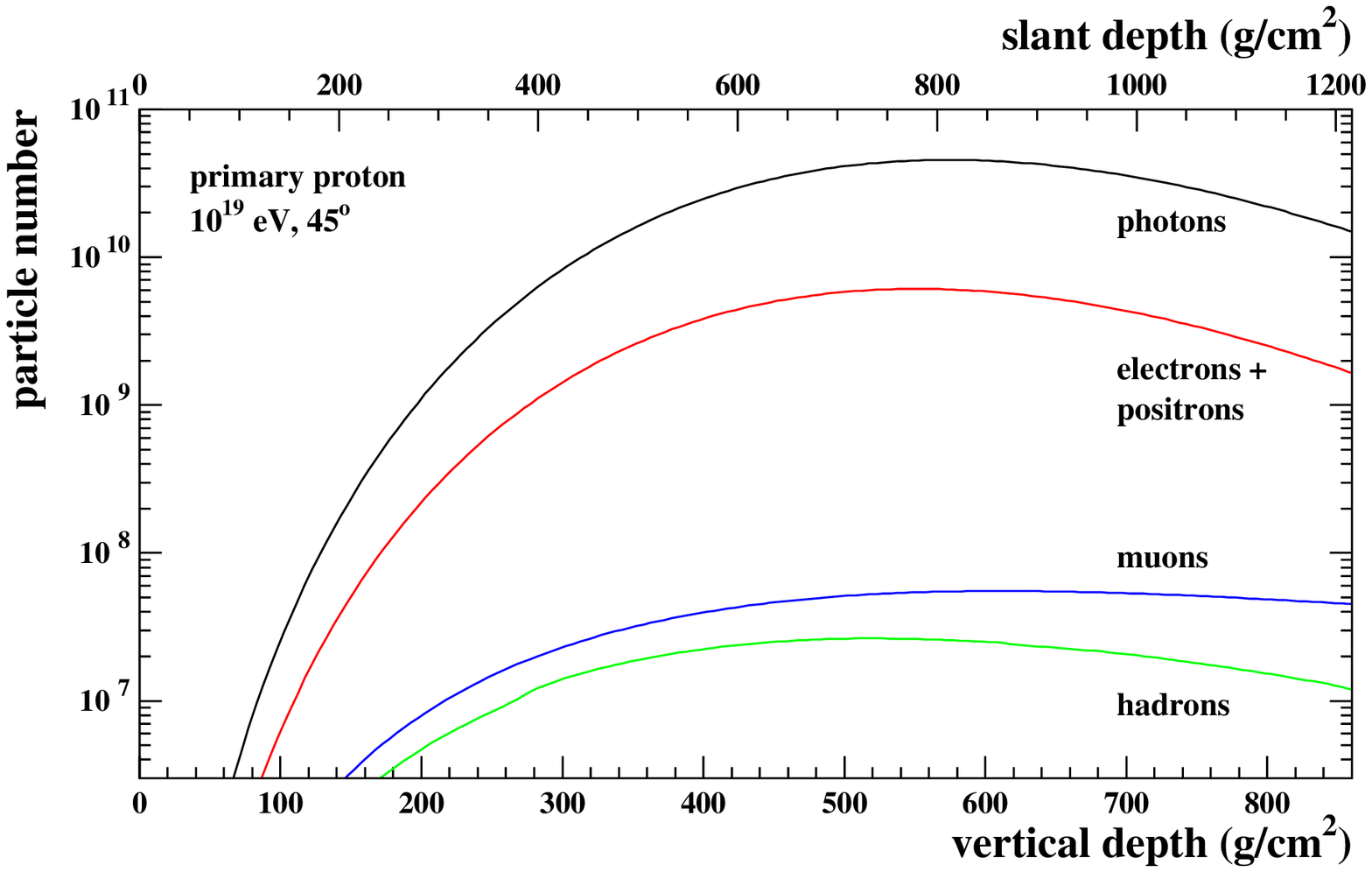}
  \caption{Simulated proton-induced extensive air shower with E$_0$ = 10$^{19}$~eV.
   Left: The overall deposited energy versus atmospheric depth is shown for the vertical
  case. The contributions from the 
  electromagnetic, muonic, and hadronic part of air showers are separated. The contributions
  labeled by 'cut' are due to simulation limitations but can be added to the
  local energy deposit~\cite{keilhauer2004a}.
   Right: The particle numbers of major contributions vs. atmospheric depth is
   shown. The shower inclination angle is 45$^\circ$, therefore the slant depth is 
   indicated at the top of the frame~\cite{risse}.}
\end{minipage}
\end{figure}

The overall longitudinal shower development, as simulated with CORSIKA\footnote{\textbf{CO}smic
\textbf{R}ay \textbf{SI}mulations for \textbf{K}ascade and \textbf{A}uger} \cite{heck}, 
is shown in Fig~\ref{fig:longidev}, where a proton-induced shower with
10$^{19}$~eV is
plotted. On the left side, the development in terms of energy deposit of a vertical
shower is displayed and on the right side, the particle number for
a 45$^\circ$ inclined shower. 

The position of the shower maximum is a good indicator for the properties of the
primary particle. The particle number at shower maximum is proportional to the
primary energy $E_0$ and the atmospheric depth of the position of shower maximum
is proportional to $ln(E_0/A)$, where $A$ is the mass of the primary.

\begin{figure}
  \label{fig:missingE}
  \includegraphics[width=.6\textwidth]{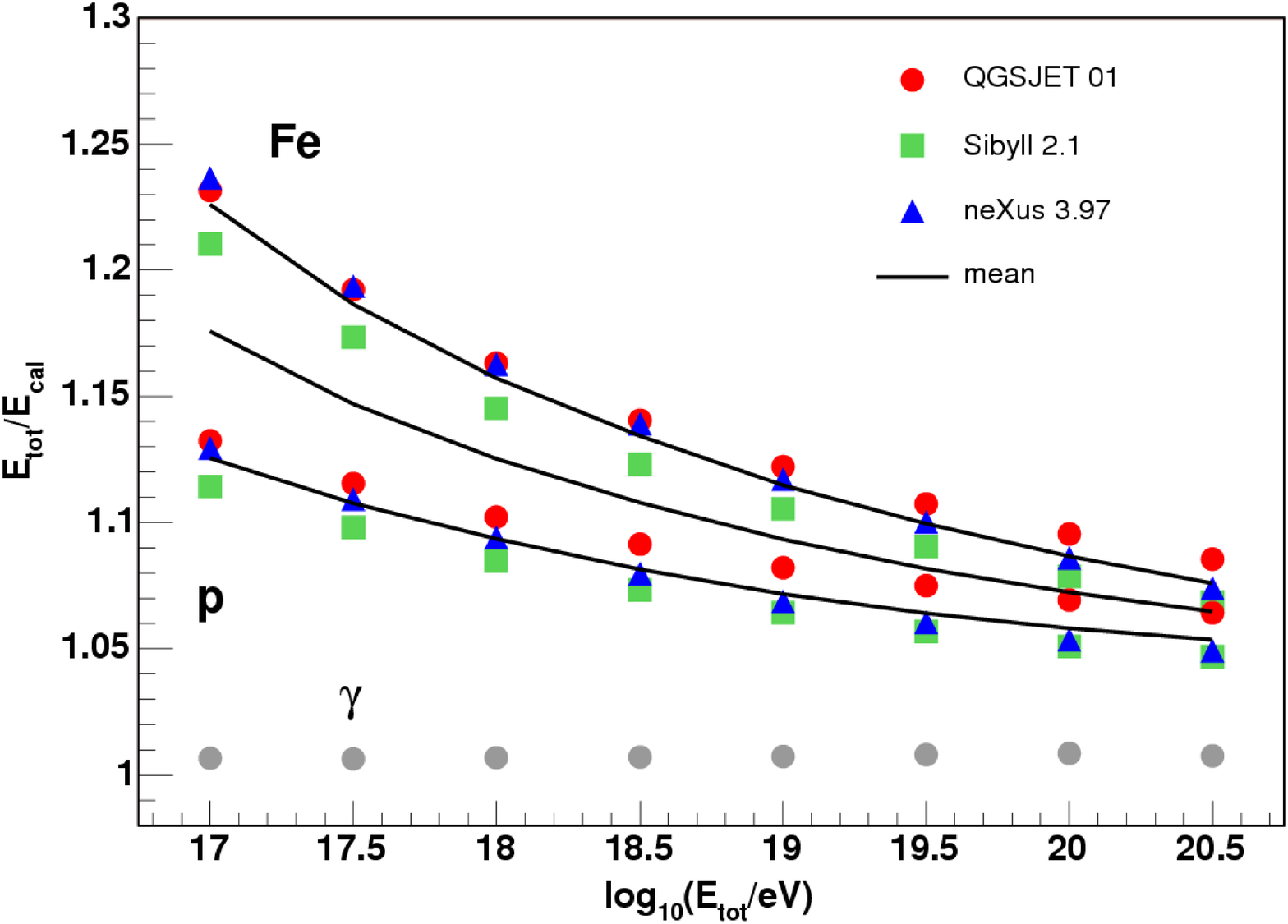}
  \caption{Correction factor for the energy which is not detected by 
  fluorescence telescopes in dependence on primary energy and type of air shower. 
  The missing energy is related to mainly neutrinos and muons as secondary particles in the 
  air shower cascade~\cite{pierog}.}
\end{figure}

Using the fluorescence technique, the energy deposit of an air shower is measured
by the amount of emitted fluorescence light. However, integrating the whole energy
deposit will not result in the total primary energy. The energy deposit is only 
that of the charged particles but the energy carried away by neutrinos is not 
detected. Therefore,
one has to perform a correction to obtain the total primary energy, which depends
on the energy and type of cosmic rays, see Fig.~\ref{fig:missingE}. 

\section{Calibration Systems}

The atmosphere is an integral part of any detector system using extensive air showers. The
longitudinal and lateral development is affected by local atmospheric conditions such as
layering of air density and temperature. The optical signal of
extensive air showers is even more influenced by the atmosphere, in the
emission and transmission towards the detector.

\begin{figure}
\label{fig:max}
\begin{minipage}[c]{.49\linewidth}
  \includegraphics[width=1.\textwidth]{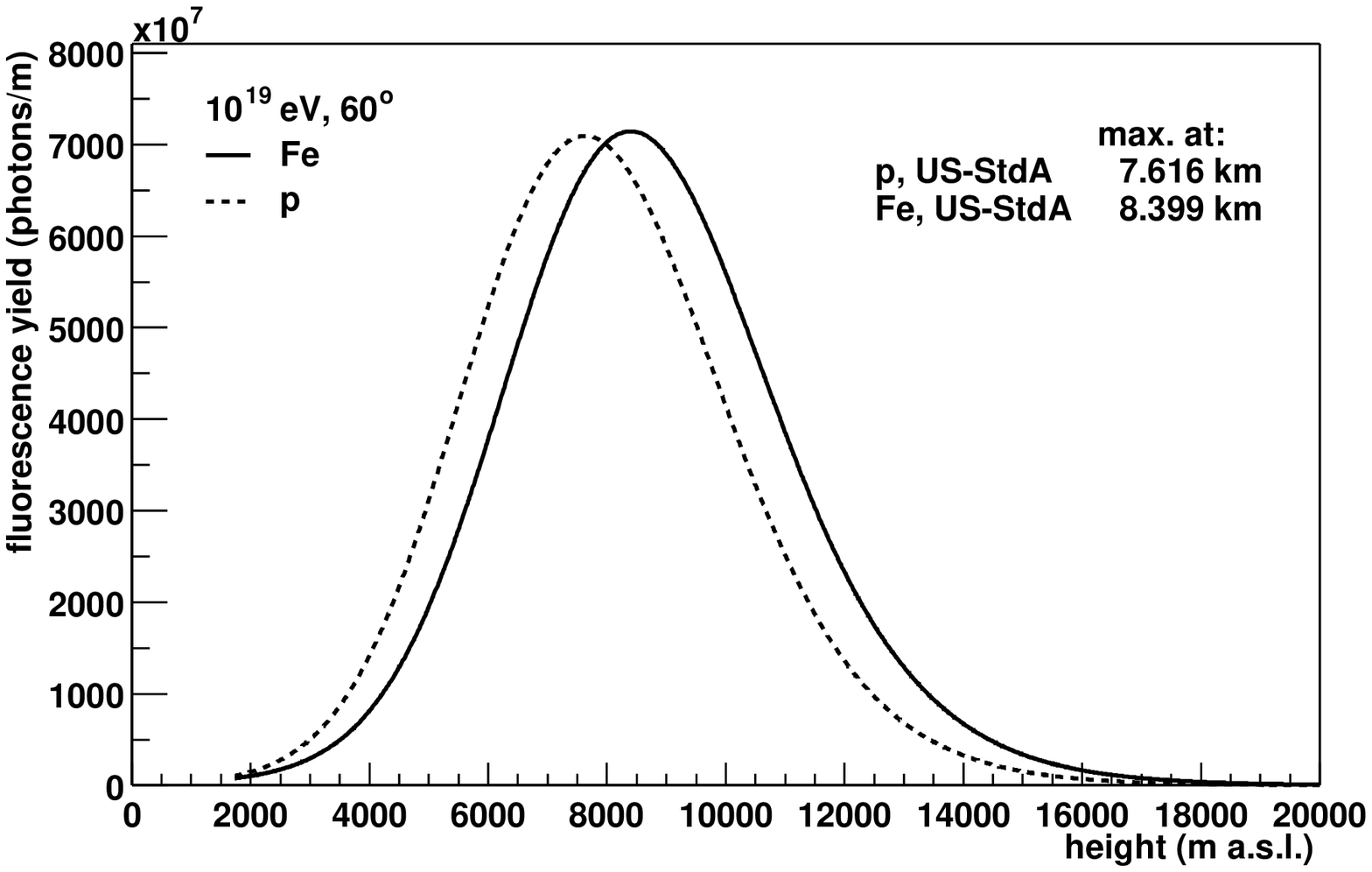}
  \caption{  }
\end{minipage}\hfill
\begin{minipage}[c]{.49\linewidth}
  \includegraphics[width=1.\textwidth]{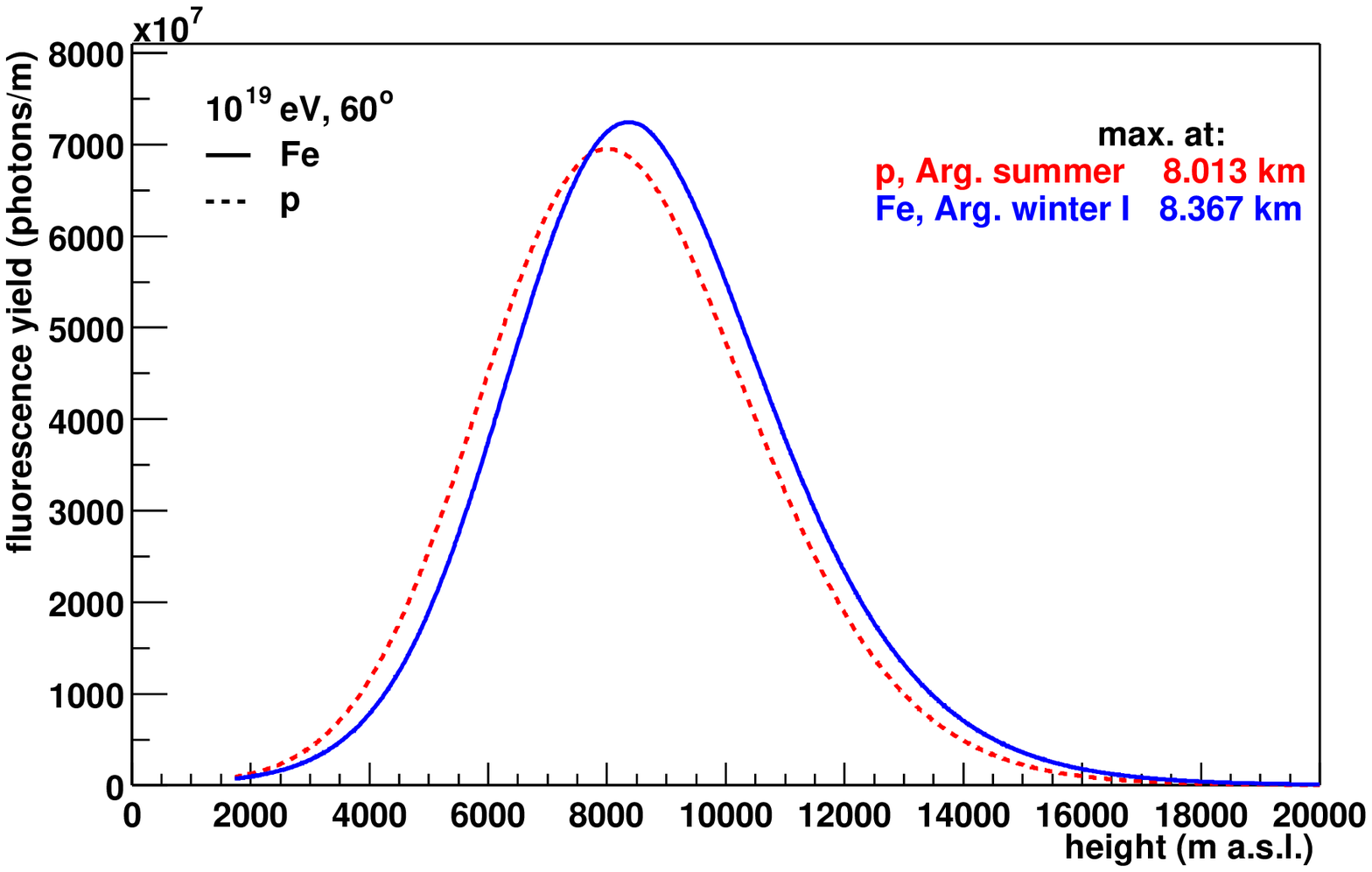}
  \caption{Left: Fluorescence yield profiles for p- and Fe-induced EAS in 
   the US Standard Atmosphere with E$_0 = 10^{19}$~eV and 60$^\circ$
   inclination. The fluorescence yield is the sum of all emitted photons
   between 300 and 400~nm~\cite{keilhauer2004a}.
   Right: Fluorescence yield profiles for p-induced EAS in Argentine summer 
   and Fe-induced EAS in Argentine winter both with E$_0 = 10^{19}$~eV and 60$^\circ$
   inclination. The fluorescence yield is the sum of all emitted photons
   between 300 and 400~nm~\cite{keilhauer2004a}.}
\end{minipage}
\end{figure}

Several calibration systems have been installed at the site of the 
Pierre Auger Observatory in Argentina. The \emph{Central Laser Facility}~\cite{CLF} and
the \emph{Lidar stations} at each telescope station~\cite{lidar} are detecting the 
optical transmission conditions of the atmosphere. The absolute \emph{drum} calibration
determines the detector response~\cite{drum}. Atmospheric variables like air temperature,
pressure, and humidity are obtained by meteorological radio
soundings~\cite{keilhauer_icrc}. From these data, the altitude and time dependent
air density can be derived which is important for the reconstruction of the 
longitudinal air shower profiles of the calorimetric measurement using air 
fluorescence emission. Varying atmospheric conditions affect the conversion of
atmospheric depth, as the characteristic quantity for the air shower development,
to geometrical altitude, as reconstructed from fluorescence telescope data. Systematic 
variations of the position of shower maximum as seen by the telescopes due to 
changing air density profiles have been found~\cite{keilhauer2004}. In Fig.~\ref{fig:max},
the position of shower maximum for simulated proton- and iron-induced showers are
shown on the left side for conditions like in the US Standard atmosphere~\cite{USStdA}.
On the right side, the same simulated showers are plotted for the case that the 
proton shower would develop in the Argentine summer atmosphere and the iron shower
in winter conditions. The separation between the positions of maximum in
the US Standard atmosphere as seen by the fluorescence telescopes is
significantly reduced.

The uncertainties on the reconstructed energy of air showers from the fluorescence
measurement are still quite large. Currently, known uncertainties for the
signals in the PMT camera are 5~\% for the light collection and 12~\% from
the absolute detector calibration~\cite{FD_ICRC}. However, the latter will
improve in the near future. For the reconstructed photons at the telescope, 
the Auger experiment has to cope with 2~\% uncertainty in the geometry reconstruction
for hybrid events and with 10~\% from aerosols in the atmosphere. The correction for
the missing energy yields 3~\% uncertainty and the varying air density profiles
2~\% for monthly models for the site of the Auger experiment.

\begin{figure}
\label{fig:flyield}
  \includegraphics[width=.7\textwidth]{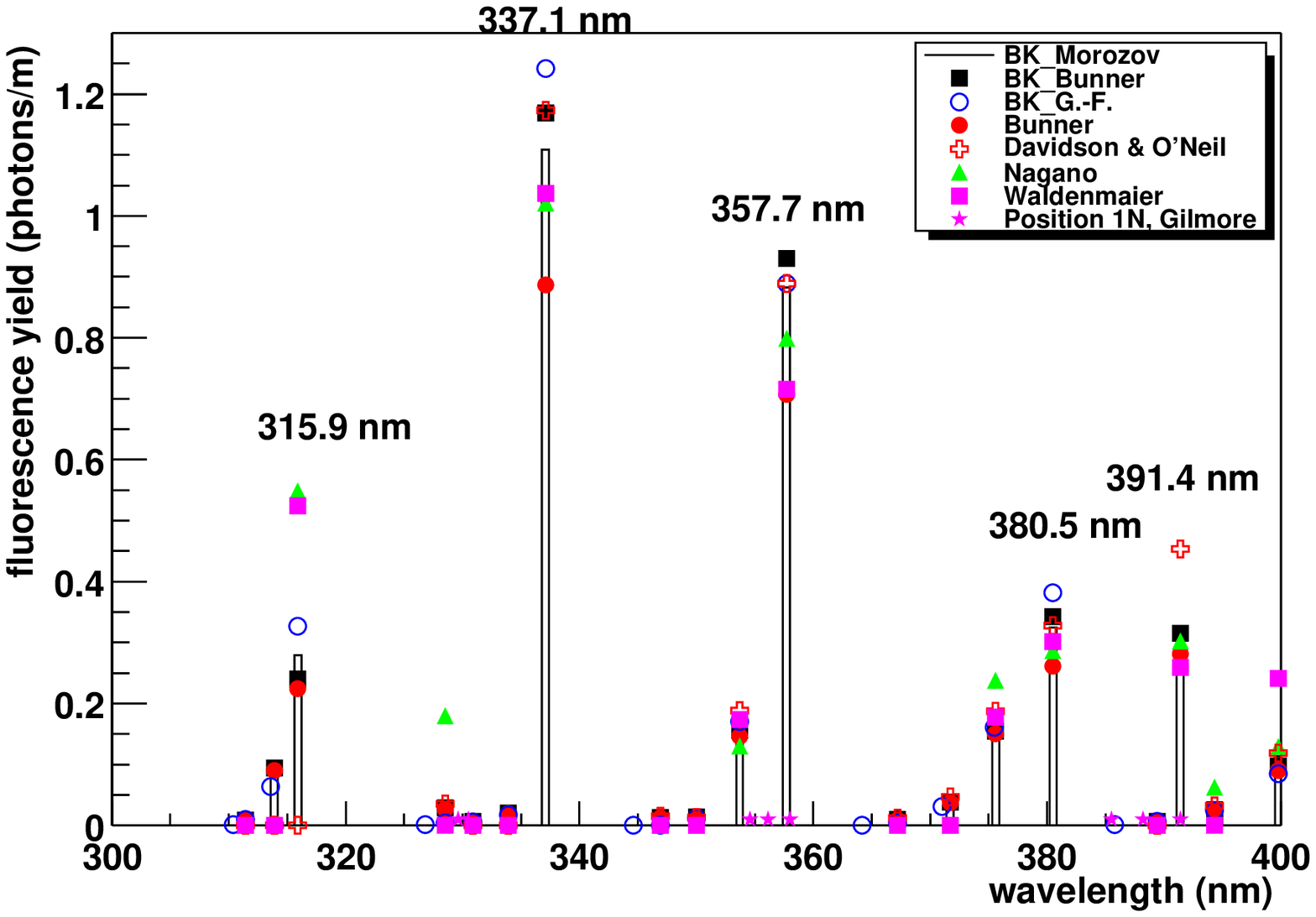}
  \caption{Fluorescence yield spectra of several calculations and measurements
   for 0.85~MeV electrons as exciting particles. The bars indicate the preferred
   calculation by the authors of the referenced article~\cite{keilhauer2006}. All
   calculations by the authors are labeled with ``BK\_\emph{name}'', where \emph{name}
   indicates the authors of the input parameters.}
\end{figure}

A very important quantity for the energy reconstruction is the fluorescence 
yield. The emitted light is mainly coming from 18 strong emission bands in
the 2P system of nitrogen molecules for the wavelength range between 300 
and 400~nm and one emission band in the 1N system. It is assumed that the 
fluorescence yield is proportional to the local energy deposit of an
air shower
\begin{eqnarray}
Fl.Yield_{\lambda} = \epsilon_\lambda (p,T)\cdot \frac{\lambda}{hc}\cdot
\frac{dE}{dX}\cdot \varrho_{air} \biggl[\frac{\rm{photons}}{\rm{m}}\biggr],
\end{eqnarray}
where $\epsilon_\lambda (p,T)$ is the fluorescence efficiency in dependence on
air pressure and temperature. Strong efforts in improving laboratory 
measurements of the main variables of the fluorescence emission are ongoing
as well as of theoretical descriptions of the molecular processes. A comparison
of current experimental data and calculations for the fluorescence yield
is given in Fig.~\ref{fig:flyield}~\cite{keilhauer2006}.
The yield is given for a 0.85~MeV electron as the exciting particle in US Standard atmosphere 
conditions at sea level. The uncertainties for individual
wavelengths are large. For the overall wavelength region, some of the
discrepancies cancel out and the uncertainty for the total fluorescence yield between 
300 and 400~nm is about 15~\% for air shower applications. However, the wavelength-dependent
light transmission coefficients and detector response necessitate a good knowledge
of the fluorescence yield of the individual emission bands. A world-wide 
cooperation of several groups will improve 
the accuracy of the most important variables linked to nitrogen fluorescence
emission in air in the near future~\cite{fl}.

\section{Cosmic Ray Energy Spectrum}

\begin{figure}
\label{fig:spectrum}
  \includegraphics[width=.7\textwidth]{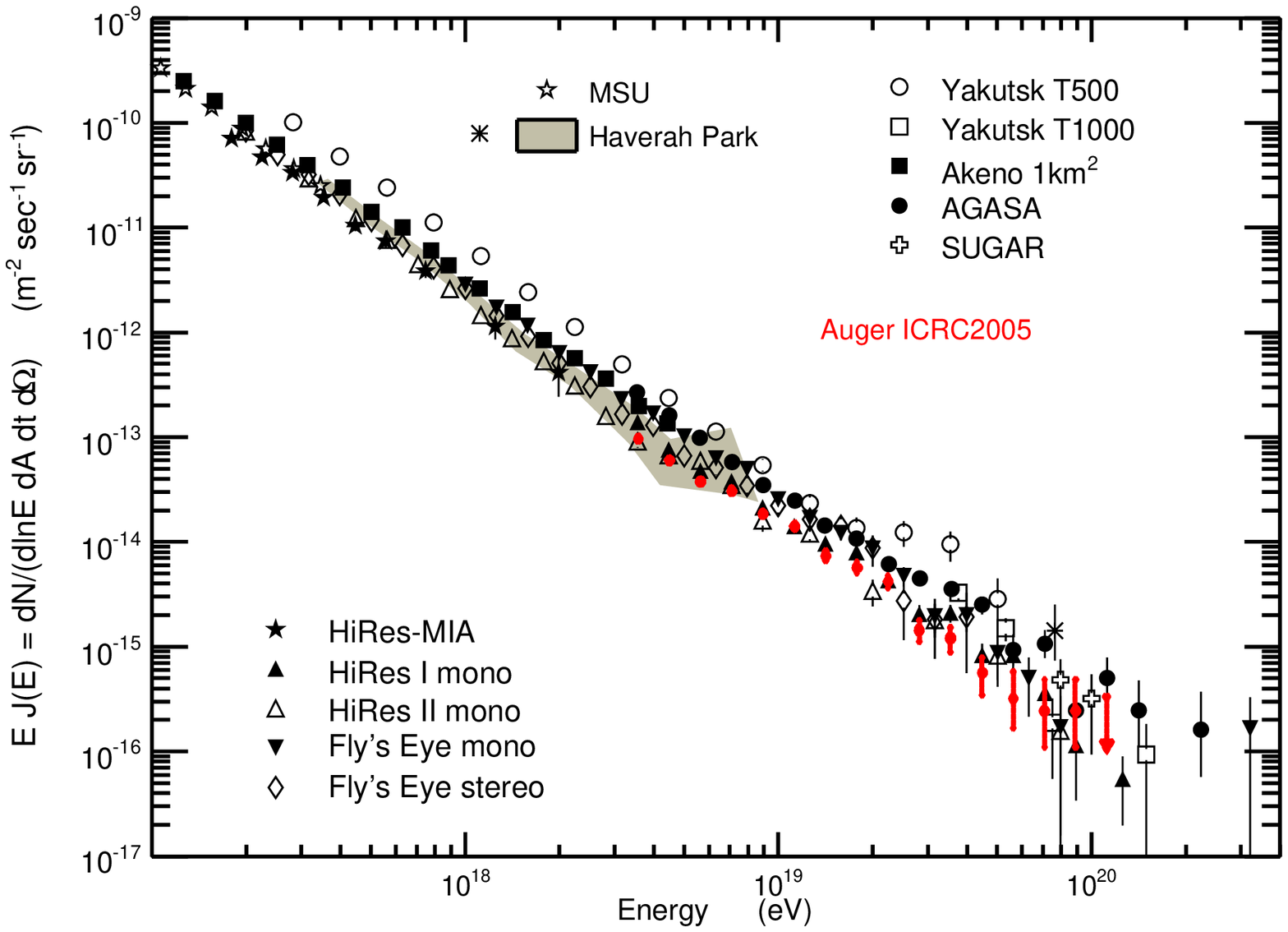}
  \caption{Primary cosmic ray flux~\cite{engel2006}. Shown is a 
   selection of previous measurements~\cite{engelklages} together with the data from the 
   Pierre Auger Observatory as presented at the ICRC 2005~\cite{sommers}.}
\end{figure}

Based on the data taken with the Pierre Auger Observatory between January 2004
and July 2005, a cosmic ray energy spectrum could be derived~\cite{sommers}. All events with
zenith angle between 0$^\circ$ and 60$^\circ$ have been taken into account.
The \emph{time average area} of the continuously growing array has been 
660~km$^2$ which is about 22~\% of the final size of the southern part of the
experiment. Applying several quality cuts, a full efficiency could be reached
above 3~EeV and 3525 events above 10$^{18.5}$~eV could be collected by the
surface array. The analysis has been performed using the following procedure.
Since the surface array could collect much more event data, this higher statistics
should be used. The zenith angle correction for these data is done with the
\emph{constant intensity cut} method~\cite{sommers}. Without fluorescence data, the 
energy reconstruction of the surface array events would depend on air shower 
simulation with its considerable
uncertainties arising from different hadronic interaction models. Therefore, 
the energy is determined from the sub-sample of hybrid events. An energy
conversion factor is derived by comparing the reconstructed energy of
hybrid events, based on the calorimetric energy measurement with the fluorescence
telescopes, with the surface detector signal at 1000~m from the shower
core. Finally, the exposure of the growing array is calculated. The estimated 
energy spectrum can be seen in Fig.~\ref{fig:spectrum} together with previous
measurements.

Statistics is still limited for the first data taking period and systematic
uncertainties are quite large. So, based on that data, no decision can be made
about the characteristics of the spectrum in the GZK cutoff region. Whether there is a systematic 
difference in the energy reconstruction for the fluorescence method and the 
surface detector method will be subject of detailed investigations. 
The Pierre Auger Observatory with its hybrid technique
intends to resolve this uncertainty together with the investigations
of the fluorescence processes. 

\begin{theacknowledgments}
  I would like to thank the organizers of the CALOR 2006 conference
  in Chicago for the invitation and the financial support. I gratefully
  acknowledge the fruitful cooperation and discussions with my 
  colleagues from the Pierre Auger Collaboration. My research is 
  supported by the German Research Foundation (DFG) under contract
  \mbox{No.~KE 1151/1-1}. 
\end{theacknowledgments}

\end{document}